\documentstyle{article}
\begin{document}

\title{\bf Comment on 'Teleportation of two quNit entanglement:
Exploiting local resources'}
\author{ Jie Yang$^a$$\thanks{Corresponding author. Email: yangjies@sina.com}$,
Hao Yuan$^{a}$ and Zhan-jun Zhang$^{a,b}$
 \\
{\normalsize $^a$ Key Laboratory of Optoelectronic Information
Acquisition \& Manipulation of Ministry of Education of China,}\\
{\normalsize School of Physics \& Material Science, Anhui
University, Hefei 230039, China} \\
{\normalsize $^b$ Department of Physics and Center for Quantum
Information Science,}\\
{\normalsize National Cheng Kung University, Tainan 70101, Taiwan} }
\date{}
\maketitle

Recently, in Ref.[1] N. Ba An proposed a teleportation protocol, in
which the new interesting idea of exploiting local resources to do
teleportation [2] is generalized to the case of arbitrary quNits.
After carefully rededucting the formulae in Ref.[1], however, we
find some of them are wrong. As a consequence, the unitary
operations which need to be performed to reconstruct the unknown
bipartite entangled state are also incorrect. This means that in
terms of N. Ba An's protocol the teleportation of two-quNit
entanglement can not be successfully realized.  To solve this
problem, in this comment we present two different efficient methods
to modify some details of N. Ba An's protocol: (1) we correct the
wrong formulae and accordingly offer right unitary operations; (2)
we redefine some definitions such that the so-called wrong formulae
and unitary operations all maintain their original forms. After our
modifications, we believe, the revised N. Ba An's protocol can work
successfully.

{\bf Method 1:} \ For convenience, throughout this method all the
definitions, such as the two-qutrit(quNit) maximally entangled (ME)
pairs $\{|\Phi_{mn}\rangle\}$, the control change gate, the rotated
basis, etc, are all same as those in the Ref.[1].

Along the line of N. Ba An's work [1], we also first consider the
case of qutrits. In this case, the whole teleportation process is
as follows: (i) Alice has two qutrits 1 and 2 in an entangled
state
$|\Psi\rangle_{12}=(\alpha|00\rangle+\beta|11\rangle+\gamma|22\rangle)_{12}$.
Besides, she also shares a single two-qutrit maximally entangled
pair in the state $|\Phi_{00}\rangle_{34}$ with Bob (say, Alice
has the qutrit 3 while Bob the qutrit 4). (ii) Alice performs a
joint measurement on the 1-3 system in the basis
$\{|\Phi_{mn}\rangle_{13}\} $ and measures qutrit 2 in a rotated
basis.  (iii) After her measurements, Alice tells Bob her two
measurement results via classical channel. (iv) Bob introduces an
ancillary  qutrit 5 and performs the control change gate $C_{45}$
on the 4-5 system. (v) Conditioned on the classical message from
Alice, Bob performs an appropriate unitary operation $U$ on the
4-5 system to reconstruct the state $|\Psi\rangle$.

In Ref[1], the total 1-2-3-4 system state in terms of  $\{
|\Phi_{mn}\rangle_{13}\}$ is written as its formula 2, that is,
\begin{eqnarray}
\frac{1}{3}&[&
(\alpha|00\rangle+\beta|11\rangle+\gamma|22\rangle)_{24}|\Phi_{00}\rangle_{13}
+(\alpha|01\rangle+\beta|12\rangle+\gamma|20\rangle)_{24}|\Phi_{01}\rangle_{13} \nonumber\\
&+&(\alpha|02\rangle+\beta|10\rangle+\gamma|21\rangle)_{24}|\Phi_{02}\rangle_{13}
+(\alpha|00\rangle+e^{-\frac{2\pi i}{3}}
\beta|11\rangle +e^{\frac{2\pi i}{3}}\gamma|22\rangle)_{24}|\Phi_{10}\rangle_{13}  \nonumber \\
&+&(\alpha|01\rangle+e^{-\frac{2\pi i}{3}}\beta|12\rangle
+e^{\frac{2\pi i}{3}}\gamma|20\rangle)_{24} |\Phi_{11}\rangle_{13}
+(\alpha|02\rangle+e^{-\frac{2\pi i}{3}}\beta|10\rangle
+e^{\frac{2\pi
i}{3}}\gamma|21\rangle)_{24} |\Phi_{12}\rangle_{13}   \nonumber \\
&+&(\alpha|00\rangle+e^{\frac{2\pi i}{3}}\beta|11\rangle
+e^{-\frac{2\pi i}{3}}\gamma|22\rangle)_{24} |\Phi_{20}\rangle_{13}
+(\alpha|01\rangle+e^{\frac{2\pi i}{3}}\beta|12\rangle
+e^{-\frac{2\pi
i}{3}}\gamma|20\rangle)_{24} |\Phi_{21}\rangle_{13}   \nonumber \\
&+&(\alpha|02\rangle+e^{\frac{2\pi i}{3}}\beta|10\rangle
+e^{-\frac{2\pi i}{3}}\gamma|21\rangle)_{24} |\Phi_{22}\rangle_{13}
 ].
\end{eqnarray}
After our careful redeductions, however, we find N. Ba An's this
expression is wrong. The correct one should be
\begin{eqnarray}
\frac{1}{3}&[&
(\alpha|00\rangle+\beta|11\rangle+\gamma|22\rangle)_{24}|\Phi_{00}\rangle_{13}
+(\alpha|01\rangle+\beta|12\rangle+\gamma|20\rangle)_{24}|\Phi_{01}\rangle_{13} \nonumber\\
&+&(\alpha|02\rangle+\beta|10\rangle+\gamma|21\rangle)_{24}|\Phi_{02}\rangle_{13}
+(e^{\frac{2\pi i}{3}}\alpha|00\rangle+e^{-\frac{2\pi i}{3}}
\beta|11\rangle +\gamma|22\rangle)_{24}|\Phi_{10}\rangle_{13}  \nonumber \\
&+&(e^{-\frac{2\pi i}{3}}\alpha|01\rangle+\beta|12\rangle
+e^{\frac{2\pi i}{3}}\gamma|20\rangle)_{24} |\Phi_{11}\rangle_{13}
+(\alpha|02\rangle+e^{\frac{2\pi i}{3}}\beta|10\rangle
+e^{-\frac{2\pi
i}{3}}\gamma|21\rangle)_{24} |\Phi_{12}\rangle_{13}   \nonumber \\
&+&(e^{-\frac{2\pi i}{3}}\alpha|00\rangle+e^{\frac{2\pi
i}{3}}\beta|11\rangle +\gamma|22\rangle)_{24}
|\Phi_{20}\rangle_{13} +(e^{\frac{2\pi
i}{3}}\alpha|01\rangle+\beta|12\rangle +e^{-\frac{2\pi
i}{3}}\gamma|20\rangle)_{24} |\Phi_{21}\rangle_{13}   \nonumber \\
&+&(\alpha|02\rangle+e^{-\frac{2\pi i}{3}}\beta|10\rangle
+e^{\frac{2\pi i}{3}}\gamma|21\rangle)_{24} |\Phi_{22}\rangle_{13}
 ].
\end{eqnarray}
Because of the mistake mentioned just, obviously, in Ref.[1] all
those formulae which are correlated to the formula 2 are also wrong.
Below we will simply give the correct formulae and point out the
correspondence between the present correct formulae and the wrong
ones in Ref.[1]. The state of 1-2-3-4-5 system, which is wrongly
expressed as the formula 4 in Ref.[1], should be written as
\begin{eqnarray}
\frac{1}{3}&[&
(\alpha|000\rangle+\beta|111\rangle+\gamma|222\rangle)_{245}|\Phi_{00}\rangle_{13}
+(\alpha|011\rangle+\beta|122\rangle+\gamma|200\rangle)_{245}|\Phi_{01}\rangle_{13} \nonumber\\
&+&(\alpha|022\rangle+\beta|100\rangle+\gamma|211\rangle)_{245}|\Phi_{02}\rangle_{13}
+(e^{\frac{2\pi i}{3}}\alpha|000\rangle+e^{-\frac{2\pi i}{3}}
\beta|111\rangle +\gamma|222\rangle)_{245}|\Phi_{10}\rangle_{13}  \nonumber \\
&+&(e^{-\frac{2\pi i}{3}}\alpha|011\rangle+\beta|122\rangle
+e^{\frac{2\pi i}{3}}\gamma|200\rangle)_{245}
|\Phi_{11}\rangle_{13} +(\alpha|022\rangle+e^{\frac{2\pi
i}{3}}\beta|100\rangle +e^{-\frac{2\pi
i}{3}}\gamma|211\rangle)_{245} |\Phi_{12}\rangle_{13}   \nonumber \\
&+&(e^{-\frac{2\pi i}{3}}\alpha|000\rangle+e^{\frac{2\pi
i}{3}}\beta|111\rangle +\gamma|222\rangle)_{245}
|\Phi_{20}\rangle_{13} +(e^{\frac{2\pi
i}{3}}\alpha|011\rangle+\beta|122\rangle +e^{-\frac{2\pi
i}{3}}\gamma|200\rangle)_{245} |\Phi_{21}\rangle_{13}   \nonumber \\
&+&(\alpha|022\rangle+e^{-\frac{2\pi i}{3}}\beta|100\rangle
+e^{\frac{2\pi i}{3}}\gamma|211\rangle)_{245}
|\Phi_{22}\rangle_{13}
 ].
\end{eqnarray}
In terms of a rotated basis for qutrit 2, the total state is
reexpressed as
\begin{eqnarray}
\frac{1}{3\sqrt{3}}&\{&|\Psi ^0 _0 \rangle_{45}[|\tilde {0}\rangle_2
|\Phi_{00}\rangle_{13}+e^{\frac{2\pi i}{3}}|\tilde {1}\rangle_2
|\Phi_{10}\rangle_{13}+e^{-\frac{2\pi i}{3}}|\tilde {2}\rangle_2
|\Phi_{20}\rangle_{13}] \nonumber \\
&+& |\Psi ^1 _0 \rangle_{45}[e^{-\frac{2\pi
i}{3}}|\tilde{0}\rangle_2 |\Phi_{20}\rangle_{13}+|\tilde
{1}\rangle_2 |\Phi_{00}\rangle_{13}+e^{\frac{2\pi i}{3}}|\tilde
{2}\rangle_2
|\Phi_{10}\rangle_{13}] \nonumber \\
&+&|\Psi ^2 _0 \rangle_{45}[e^{\frac{2\pi i}{3}}|\tilde {0}\rangle_2
|\Phi_{10}\rangle_{13}+e^{-\frac{2\pi i}{3}} |\tilde {1}\rangle_2
|\Phi_{20}\rangle_{13}+|\tilde {2}\rangle_2 |\Phi_{00}\rangle_{13}]
\nonumber \\
&+&|\Psi ^0 _1 \rangle_{45}[|\tilde {0}\rangle_2
|\Phi_{01}\rangle_{13}+e^{-\frac{2\pi i}{3}} |\tilde{1}\rangle_2
|\Phi_{11}\rangle_{13}+e^{\frac{2\pi i}{3}}|\tilde{2}\rangle_2
|\Phi_{21}\rangle_{13}]\nonumber \\
&+&|\Psi ^1 _1 \rangle_{45}[e^{\frac{2\pi i}{3}}|\tilde {0}\rangle_2
|\Phi_{21}\rangle_{13}+ |\tilde{1}\rangle_2
|\Phi_{01}\rangle_{13}+e^{-\frac{2\pi i}{3}}|\tilde{2}\rangle_2
|\Phi_{11}\rangle_{13}]\nonumber \\
&+&|\Psi ^2 _1 \rangle_{45}[e^{-\frac{2\pi i}{3}}|\tilde
{0}\rangle_2 |\Phi_{11}\rangle_{13}+e^{\frac{2\pi i}{3}}|\tilde
{1}\rangle_2 |\Phi_{21}\rangle_{13}+|\tilde{2}\rangle_2
|\Phi_{01}\rangle_{13}]\nonumber \\
&+&|\Psi ^0 _2 \rangle_{45}[|\tilde{0}\rangle_2
|\Phi_{02}\rangle_{13} + |\tilde {1}\rangle_2
|\Phi_{12}\rangle_{13}+|\tilde {2}\rangle_2 |\Phi_{22}\rangle_{13}] \nonumber \\
&+& |\Psi ^1 _2 \rangle_{45}[|\tilde{0}\rangle_2
|\Phi_{22}\rangle_{13} + |\tilde {1}\rangle_2
|\Phi_{02}\rangle_{13}+|\tilde{2}\rangle_2
|\Phi_{12}\rangle_{13}]\nonumber \\
&+&|\Psi ^2 _2 \rangle_{45}[|\tilde {0}\rangle_2
|\Phi_{12}\rangle_{13} + |\tilde {1}\rangle_2
|\Phi_{22}\rangle_{13}+|\tilde {2}\rangle_2 |\Phi_{02}\rangle_{13}]
 \}.
\end{eqnarray}
This formula corresponds to the wrong formula 5 in Ref.[1]. In the
teleportation the unitary operations which need to be performed to
reconstruct the unknown bipartite entangled state are recognized and
abstracted from the reexpression in terms of a rotated basis for
qutrit 2. Since in Ref.[1] the reexpression(i.e., the formula 5 in
Ref.[1]) is wrong, the 9 unitary operations $U_{lmn}$ abstracted
from the reexpression  and shown in Table 1 of Ref.[1] are of course
unreliable anymore. We have anew abstracted the 9 unitary operations
$U_{lmn}$ from the present formula 4 and displayed them in the
present Table 1. Comparing the present Table with that in Ref.[1],
one can easily find some differences. For an example, when Alice's
measurement is$\{2,2,0\}$, Bob should perform operation $|00\rangle
\langle00|+e^{\frac{2\pi i}{3}}|11\rangle \langle11|+e^{-\frac{2\pi
i}{3}}|22\rangle \langle22|$ on the 4-5 system according to the
table in Ref.[1], while Bob needs to do nothing according to the
present Table (See the third line of the Table). \\

\begin{center}
\noindent Table 1 Bob's operations $U_{lmn}$ performed on the 4-5
system conditioned on Alice's measurement outcome$\{l,m,n\}$

\begin{tabular}{lllllllll}\hline
$l$& &$m$& &$n$& &$U_{lmn}({\rm Ours})$& &$U_{lmn}({\rm N. Ba
An's})$
\\\hline
$0$ & & $0$ & & $0$& & $\hat I$ & &
$\hat I$\\
$1$ & & $1$ & & $0$& & $\hat I$ & &
$|00\rangle \langle00|+e^{-\frac{2\pi i}{3}}|11\rangle \langle11|+e^{\frac{2\pi i}{3}}|22\rangle \langle22|$\\
$2$ & & $2$ & & $0$& & $\hat I$ & &
$|00\rangle \langle00|+e^{\frac{2\pi i}{3}}|11\rangle \langle11|+e^{-\frac{2\pi i}{3}}|22\rangle \langle22|$\\

$0$& &$2$& &$0$& &
$|00\rangle \langle00|+e^{\frac{2\pi i}{3}}|11\rangle \langle11|+e^{-\frac{2\pi i}{3}}|22\rangle \langle22|$& &
$|00\rangle \langle00|+e^{-\frac{2\pi i}{3}}|11\rangle \langle11|+e^{\frac{2\pi i}{3}}|22\rangle \langle22|$ \\
$1$& &$0$& &$0$& &
$|00\rangle \langle00|+e^{\frac{2\pi i}{3}}|11\rangle \langle11|+e^{-\frac{2\pi i}{3}}|22\rangle \langle22|$& &
$|00\rangle \langle00|+e^{\frac{2\pi i}{3}}|11\rangle \langle11|+e^{-\frac{2\pi i}{3}}|22\rangle \langle22|$ \\
$2$& &$1$& &$0$& &
$|00\rangle \langle00|+e^{\frac{2\pi i}{3}}|11\rangle \langle11|+e^{-\frac{2\pi i}{3}}|22\rangle \langle22|$& &
$\hat I$\\

$0$& &$1$& &$0$& &
$|00\rangle \langle00|+e^{-\frac{2\pi i}{3}}|11\rangle \langle11|+e^{\frac{2\pi i}{3}}|22\rangle \langle22|$ & &
$|00\rangle \langle00|+e^{\frac{2\pi i}{3}}|11\rangle \langle11|+e^{-\frac{2\pi i}{3}}|22\rangle \langle22|$\\
$1$& &$2$& &$0$& &
$|00\rangle \langle00|+e^{-\frac{2\pi i}{3}}|11\rangle \langle11|+e^{\frac{2\pi i}{3}}|22\rangle \langle22|$ & &
$\hat I$\\
$2$& &$0$& &$0$& &
$|00\rangle \langle00|+e^{-\frac{2\pi i}{3}}|11\rangle \langle11|+e^{\frac{2\pi i}{3}}|22\rangle \langle22|$ & &
$|00\rangle \langle00|+e^{-\frac{2\pi i}{3}}|11\rangle \langle11|+e^{\frac{2\pi i}{3}}|22\rangle \langle22|$\\

$0$ & &$0$ & &$1$& &
$|00\rangle \langle11|+|11\rangle \langle22|+|22\rangle \langle00|$ & &
$|00\rangle \langle11|+|11\rangle \langle22|+|22\rangle \langle00|$ \\
$1$ & &$1$ & &$1$& &
$|00\rangle \langle11|+|11\rangle \langle22|+|22\rangle \langle00|$ & &
$|00\rangle \langle11|+e^{-\frac{2\pi i}{3}}|11\rangle \langle22|+e^{\frac{2\pi i}{3}}|22\rangle \langle00|$\\
$2$ & &$2$ & &$1$& &
$|00\rangle \langle11|+|11\rangle \langle22|+|22\rangle \langle00|$ & &
$|00\rangle \langle11|+e^{\frac{2\pi i}{3}}|11\rangle \langle22|+e^{-\frac{2\pi i}{3}}|22\rangle \langle00|$ \\

$0$& & $2$ & & $1$& &
$|00\rangle \langle11|+e^{\frac{2\pi i}{3}}|11\rangle \langle22|+e^{-\frac{2\pi i}{3}}|22\rangle \langle00|$ & &
$|00\rangle \langle11|+e^{-\frac{2\pi i}{3}}|11\rangle \langle22|+e^{\frac{2\pi i}{3}}|22\rangle \langle00|$\\
$1$& & $0$ & & $1$& &
$|00\rangle \langle11|+e^{\frac{2\pi i}{3}}|11\rangle \langle22|+e^{-\frac{2\pi i}{3}}|22\rangle \langle00|$ & &
$|00\rangle \langle11|+e^{\frac{2\pi i}{3}}|11\rangle \langle22|+e^{-\frac{2\pi i}{3}}|22\rangle \langle00|$ \\
$2$& & $1$ & & $1$& &
$|00\rangle \langle11|+e^{\frac{2\pi i}{3}}|11\rangle \langle22|+e^{-\frac{2\pi i}{3}}|22\rangle \langle00|$ & &
$|00\rangle \langle11|+|11\rangle \langle22|+|22\rangle \langle00|$ \\

$0$ & & $1$ & & $1$& &
$|00\rangle \langle11|+e^{-\frac{2\pi i}{3}}|11\rangle \langle22|+e^{\frac{2\pi i}{3}}|22\rangle \langle00|$ & &
$|00\rangle \langle11|+e^{\frac{2\pi i}{3}}|11\rangle \langle22|+e^{-\frac{2\pi i}{3}}|22\rangle \langle00|$ \\
$1$ & & $2$ & & $1$& &
$|00\rangle \langle11|+e^{-\frac{2\pi i}{3}}|11\rangle \langle22|+e^{\frac{2\pi i}{3}}|22\rangle \langle00|$ & &
$|00\rangle \langle11|+|11\rangle \langle22|+|22\rangle \langle00|$ \\
$2$ & & $0$ & & $1$& &
$|00\rangle \langle11|+e^{-\frac{2\pi i}{3}}|11\rangle \langle22|+e^{\frac{2\pi i}{3}}|22\rangle \langle00|$ & &
$|00\rangle \langle11|+e^{-\frac{2\pi i}{3}}|11\rangle \langle22|+e^{\frac{2\pi i}{3}}|22\rangle \langle00|$\\

$0$ & & $0$ & &$2$& &
$|00\rangle \langle22|+|11\rangle \langle00|+|22\rangle \langle11|$ & &
$|00\rangle \langle22|+|11\rangle \langle00|+|22\rangle \langle11|$\\
$1$ & & $1$ & &$2$& &
$|00\rangle \langle22|+|11\rangle \langle00|+|22\rangle \langle11|$ & &
$|00\rangle \langle22|+e^{-\frac{2\pi i}{3}}|11\rangle \langle00|+e^{\frac{2\pi i}{3}}|22\rangle \langle11|$\\
$2$ & & $2$ & &$2$& &
$|00\rangle \langle22|+|11\rangle \langle00|+|22\rangle \langle11|$ & &
$|00\rangle \langle22|+e^{\frac{2\pi i}{3}}|11\rangle \langle00|+e^{\frac{-2\pi i}{3}}|22\rangle \langle11|$\\

$0$ & &$2$ & &$2$& &
$|00\rangle \langle22|+e^{\frac{2\pi i}{3}}|11\rangle \langle00|+e^{\frac{-2\pi i}{3}}|22\rangle \langle11|$ & &
$|00\rangle \langle22|+e^{-\frac{2\pi i}{3}}|11\rangle \langle00|+e^{\frac{2\pi i}{3}}|22\rangle \langle11|$\\
$1$ & &$0$ & &$2$& &
$|00\rangle \langle22|+e^{\frac{2\pi i}{3}}|11\rangle \langle00|+e^{\frac{-2\pi i}{3}}|22\rangle \langle11|$ & &
$|00\rangle \langle22|+e^{\frac{2\pi i}{3}}|11\rangle \langle00|+e^{\frac{-2\pi i}{3}}|22\rangle \langle11|$\\
$2$ & &$1$ & &$2$& &
$|00\rangle \langle22|+e^{\frac{2\pi i}{3}}|11\rangle \langle00|+e^{\frac{-2\pi i}{3}}|22\rangle \langle11|$ & &
$|00\rangle \langle22|+|11\rangle \langle00|+|22\rangle \langle11|$\\

$0$ & & $1$ & & $2$& &
$|00\rangle \langle22|+e^{-\frac{2\pi i}{3}}|11\rangle \langle00|+e^{\frac{2\pi i}{3}}|22\rangle \langle11|$& &
$|00\rangle \langle22|+e^{\frac{2\pi i}{3}}|11\rangle \langle00|+e^{\frac{-2\pi i}{3}}|22\rangle \langle11|$ \\
$1$ & & $2$ & & $2$& &
$|00\rangle \langle22|+e^{-\frac{2\pi i}{3}}|11\rangle \langle00|+e^{\frac{2\pi i}{3}}|22\rangle \langle11|$& &
$|00\rangle \langle22|+|11\rangle \langle00|+|22\rangle \langle11|$ \\
$2$ & & $0$ & & $2$& &
$|00\rangle \langle22|+e^{-\frac{2\pi i}{3}}|11\rangle \langle00|+e^{\frac{2\pi i}{3}}|22\rangle \langle11|$& &
$|00\rangle \langle22|+e^{-\frac{2\pi i}{3}}|11\rangle \langle00|+e^{\frac{2\pi i}{3}}|22\rangle \langle11|$\\
\hline
\end{tabular}\\
\end{center}

\vskip 0.8cm

Now let us turn to the case of quNits. In this case, the error also
starts from the expression of the total 1-2-3-4 system state in
terms of $|\Phi_{mn}\rangle_{13}$. In Ref[1], this state is written
as
\begin{equation}
\frac{1}{N} \sum^{N-1}_{k,n,m=0} \alpha_k e^{-\frac{2\pi i
km}{N}}|\Phi_{mn}\rangle_{13} |k,[k+n]_N \rangle_{24}.
\end{equation}
After our careful rededuction, however, we believe that the correct
expression should be
\begin{equation}
\frac{1}{N} \sum^{N-1}_{k,n,m=0} \alpha_k e^{\frac{2\pi i
([n+k]_N+1)m}{N}}|\Phi_{mn}\rangle_{13} |k,[k+n]_N \rangle_{24}.
\end{equation}
Same as the qutrit case, in Ref.[1] all the following formulae
which are correlated to the total 1-2-3-4 system state in terms of
$|\Phi_{mn}\rangle_{13}$ are also wrong. The state of 2-4-5
system, which is wrongly expressed as the formula 12 in Ref.[1],
should be written as
\begin{equation}
\sum^{N-1}_{k=0} \alpha_k e^{\frac{2\pi i
([n+k]_N+1)m}{N}}|k\rangle_{2} |[k+n]_N,[k+n]_N \rangle_{45}.
\end{equation}
In terms of a rotated basis for quNit 2, this state can be
reexpressed as
\begin{equation}
\frac{1}{\sqrt N}\sum^{N-1}_{k,l=0} \alpha_k e^{\frac{2\pi i
(km+mn-kl+m)}{N}}|\tilde{l}\rangle_{2} |[k+n]_N,[k+n]_N
\rangle_{45}.
\end{equation}
If Alice measures quNit 2 and finds $ |\tilde{l}\rangle_2$, Bob's
two quNits 4 and 5 collapse into the state
\begin{equation}
|\Psi^{[l-m]_N}_n\rangle_{45}=\sum^{N-1}_{k=0} \alpha_k
e^{-\frac{2\pi i k(l-m)}{N}}|[k+n]_N,[k+n]_N \rangle_{45}.
\end{equation}
Our this corrected expression corresponds to the wrong formula 13 in
Ref.(1). From the above formula, the $N^2$ unitary operations Bob
has to perform on his 4-5 system to reconstruct the unknown
bipartite entangled state should be expressed as
\begin{equation}
U^{[l-m]_N} _n=\sum^{N-1}_{q=0} e^{\frac{2\pi i q[l-m]_N}{N}}|q,q
\rangle_{45} \langle[q+n]_N,[q+n]_N|.
\end{equation}
where $U^{[l-m]_N} _n$ is identified by ${[l-m]_N}$ and $n$. That is
to say, if Alice's measurement is $\{l,m,n\}$, the unitary operation
Bob should adopt is $U^{[l-m]_N} _n$ rather than $U^{[l+m]_N} _n$ of
formula 14 in Ref.[1].

{\bf Method 2:}\ By means of careful investigations and rigorous
redeductions, fortunately we find the problem that the original N.
Ba An's protocol can not work successfully can be solved by
redefining some initial definitions. More specifically, in the case
of quNits, the two-quNit ME pairs $|\Phi_{mn} \rangle_{XY}$ (i.e.,
the formula 9 in Ref.[1]) needs to be redefined as
\begin{equation}
|\Phi_{mn} \rangle_{XY}=\frac{1}{\sqrt {N}}\sum^{N-1}_{q=0} e^{\frac
{2\pi i qm}{N}} |q,[N-1-q-n]_N \rangle_{XY}.
\end{equation}
When $N=3$, this corresponds to the expressions of the two-qutrit ME
pairs $|\Phi_{mn}\rangle_{13}$. After this redefinition, the
so-called wrong formulae and unitary operations which are revealed
in the first method all maintain their original forms in Ref.[1],
and thus the N. Ba An's protocol can work successfully.

In conclusion, in this comment we reveal the problem that the
teleportation of two-quNit entanglement can not be successfully
achieved in terms of the original N. Ba An's protocol[1]. The
essential reason arising the problem is that there exist the
conflicts between the original definitions and some formulae in
Ref.[1]. To solve the problem we present two different efficient
methods to modify the original N. Ba An's protocol. After our
modifications, the revised N. Ba An's protocol can work
successfully.  \\

\noindent {\bf Acknowledgements}

This work is supported by the National Natural Science Foundation
of China under Grant Nos. 60677001 and 10304022, the
science-technology fund of Anhui province for outstanding youth
under Grant No.06042087, the general fund of the educational
committee of Anhui province under Grant No.2006KJ260B, and the key
fund of the ministry
of education of China under Grant No.206063. \\

\noindent {\bf References}

\noindent[1] N. Ba An, Phys. Lett. A 341 (2005) 9.

\noindent[2] M. M. Cola, M. G. A. Paris, Phys. Lett. A 337 (2005)
10.

\enddocument

\title{\bf Comment on 'Reply to Yang et al.'s comment'}
\author{ Jie Yang$\thanks{Corresponding author. Email:
yangjies@sina.com}$  and Hao Yuan
 \\
{\normalsize Key Laboratory of Optoelectronic Information
Acquisition \& Manipulation of Ministry of Education of China,}\\
{\normalsize School of Physics \& Material Science, Anhui
University, Hefei 230039, China}  }
\date{}
\maketitle
\begin{center}
This is a comment on 'Reply to Yang et al.'s comment' by N. Ba An.
\end{center}

In our recent comment[1] on 'Teleportation of two quNit
entanglement: Exploiting local resources[2]', we have revealed that,
in terms of Ba An's protocol[2] teleportation of two quNit
entanglement can not be realized at all due to some mistakes in the
protocol. Moreover, in the comment we have also presented two
efficient methods to revise Ba An's protocol such that the revised
protocol can work.

In Ba An's reply[3] to our comment, he claimed that our first method
is complicated and unnecessary for he thought the mistakes in his
protocol can be corrected simply by exchanging the subscripts 3 and
4. Hence, he suggested to publish an Erratum to correct the
mistakes. But after our careful checks, we think his method of
exchanging the subscripts 3 and 4 is still invalid at all, i.e., the
mistakes in Ba An's protocol[3] can not be corrected and accordingly
the teleportation of two quNit entanglement can not be realized.
Therefore, we strongly disagree his suggestion of publishing only an
Erratum with thanks to us to correct the mistakes in his paper[2].

It should be emphasized that our first method is indeed valid. In
contrast, his so-called simple method of exchanging the subscripts
can not make his original protocol valid at all. Incidentally, Ba
An[3] has already confirmed the validity of our second method and
just simply repeated our this method in his reply. \\

\noindent {\bf References}

\noindent[1] J. Yang, H. Yuan and Z. J. Zhang, quant-ph/0611184.

\noindent[2] N. Ba An, Phys. Lett. A 341 (2005) 9.

\noindent[3] N. Ba An, quant-ph/0611207.

\enddocument